\documentclass[conference]{IEEEtran}
\usepackage{graphics}
\usepackage{graphicx}
\usepackage{epsfig,amssymb}
\usepackage{epstopdf}
\usepackage{amsmath}
\usepackage{breqn}
\usepackage{mathrsfs}
\usepackage{array}
\usepackage{amsmath, latexsym, amsfonts, amssymb}
\usepackage[mathscr]{eucal}
\usepackage{array, tabularx}
\usepackage[table]{xcolor}
\usepackage{multirow}
\usepackage{epsfig}
\usepackage{algorithm}
\usepackage[noend]{algpseudocode}
\usepackage{cite}
\usepackage[numbers,sort & compress]{natbib}

\usepackage{tikz}
\usepackage{url}
\usepackage{mathtools}
\usepackage{psfragx}
\usepackage{xcolor}
\usepackage{authblk}
\usepackage{balance}

\newcommand{\paren}[1]{\left(#1\right)}
\newcommand{\sqparen}[1]{\left[#1\right]}
\newcommand{\brparen}[1]{\left\{#1\right\}}

\newcommand{\defeq}{\ensuremath{\triangleq}} 
\newcommand{\e}[1]{\ensuremath{{\rm e}^{#1}}} 

\renewcommand{\vec}[1]{\ensuremath{\boldsymbol{#1}}} 
\newcommand{\ie}{\ensuremath{{\text{\em i.e.}}}}

\newtheorem{theorem}{Theorem}
\newtheorem{lemma}{Lemma}

\newtheorem{corollary}{Corollary}

\IEEEoverridecommandlockouts

\begin{document}
\title{Wireless Energy Transfer to a Pair of Energy Receivers using Signal Strength Feedback}

%

\author[$ \dag $]{Chanaka Singhabahu}
\author[$ \dag $]{Tharaka Samarasinghe}
\author[$ \ast \ddag $]{Samith Abeywickrama}
\author[$ \ddag  $]{Chau Yuen}
\affil[$ \dag $]{University of Moratuwa, Sri Lanka}
\affil[$ \ddag $]{Singapore University of Technology and Design,  Singapore}
\affil[$ \ast $]{National University of Singapore, Singapore }
	
\affil[$  $ ]{Email: \textit {mails.chanaka@gmail.com, tharaka@ent.mrt.ac.lk, tharindu@mymail.sutd.edu.sg,  yuenchau@sutd.edu.sg } }

\date{}
\bibliographystyle{ieeetr}
\maketitle
\vspace{-1.5cm}
\begin{abstract}
This paper focuses on wireless energy transfer (WET) to a pair of low complex energy receivers (ER), by only utilizing received signal strength indicator (RSSI) values that are fed back from the ERs to the energy transmitter (ET). Selecting the beamformer that maximizes the total average energy transfer between the ET and the ERs, while satisfying a minimum harvested energy criterion at each ER, is studied. This is a non-convex constrained optimization problem which is difficult to solve analytically. Also, any analytical solution to the problem should only consists of parameters that the ET knows, or the ET can estimate, as utilizing only RSSI feedback values for channel estimation prohibits estimating some channel parameters. Thus, the paper focuses on obtaining a suboptimal solution analytically. It is proven that if the channels between the ET and the ERs satisfy a certain sufficient condition, this solution is in fact optimal. Simulations show that the optimality gap is negligibly small as well. Insights into a system with more than two ERs are also presented. To this end, it is highlighted that if the number of ERs is large enough, it is possible to always find a pair of ERs satisfying the sufficient condition, and hence, a pairwise scheduling policy that does not violate optimality can be used for the WET.\end{abstract}

\section{Introduction}
Wireless energy transfer (WET) using radio frequency (RF) signals has emerged as a promising technology to deliver energy to remotely located electronic devices over the air interface \cite{survey}. Among several techniques that facilitate WET, energy beamforming, that utilizes multiple antennas at the energy transmitter (ET), has attracted considerable research interest \cite{power1,multi_ant,multi_mimo,6623072,7060673}. It is well known that the availability of channel state information (CSI) is vital for the effectiveness of energy beamforming \cite{6568923}. This paper studies how the WET through energy beamforming can still be performed effectively, when perfect CSI is unavailable at the ET.   

CSI is usually obtained through a channel estimation process at the energy receiver (ER) in the training stage. However, due to tight energy constraints as well as hardware limitations, performing channel estimation at low complex ERs may become infeasible. Hence, we avoid conventional pilot based techniques, where the channel estimation or signal-to-interference-plus-noise
(SINR) ratio calculation is done at the receiver \cite{Tse02,hassibi,ts_on_the_outage,ts_modandanalysis}, and focuses on channel estimation at the transmitter \cite{one_bit,recip3,rssi_work, our_rssi, our_tsp}. To this end, \cite{one_bit} proposes a method to estimate the channel, at the ET, using a one-bit feedback algorithm, and \cite{recip3} exploits channel reciprocity for channel learning. \cite{rssi_work, our_rssi, our_tsp} propose received signal strength indicator (RSSI) based channel learning, and out of them, \cite{our_rssi, our_tsp} can be considered to be the most related to our work. \cite{our_rssi, our_tsp} propose a methodology for estimating the phase values of the channels between an ER and each antenna of a multi-antenna ET, using RSSI feedback values. These estimates are utilized to employ equal gain transmit (EGT) beamforming for WET, and the authors go on to show that the proposed low complex methodology can be in fact implemented on hardware. Note that RSSI values are already available in many receivers, and no significant signal processing is needed to obtain them. In this paper, we take a gradual step by introducing two ERs, and focus on the fundamental question on how the beamformer can be set at the ET in such a scenario, by only utilizing the RSSI feedback values from the ERs.

To this end, we set the beamformer at the ET with a goal of maximizing the total average energy transfer between the ET and the ERs. However, we ensure that a minimum harvested energy criterion for each ER is fulfilled, so that they can stay operational. In this way, we guarantee some fairness in the WET, while being conscious about the overall performance. The problem is a non-trivial and non-convex constrained optimization problem, and not having access to perfect CSI at the ET also creates a bottleneck.  This is because any analytical solution that we obtain should only be a function of channel parameters that the ET knows, or the ET can estimate using the RSSI feedback values. Obviously, it will be impossible for the ET to estimate all channel parameters by only utilizing the RSSI feedback values. Hence, we focus on analytically obtaining a suboptimal solution, that can be practically implemented according to our setup. However, although being suboptimal, simulations justify that the optimality gap is negligibly small. To this end, we show that if the channels between the ET and the ERs satisfy a certain sufficient condition, the solution will be in fact optimal. Results also shed light into a system having more than two ERs. It can be seen that we can always find a pair of ERs satisfying the aforementioned sufficient condition if the number of ERs in the network is large. Hence, we can adhere to a pairwise scheduling policy without violating optimality. 

The paper is organized as follows. In Section \ref{Section:System model}, we introduce the system model and set up the optimization problem. In Section \ref{Section:Solving Optimization}, we discuss how the optimization problem can be solved, considering different scenarios. In Section \ref{Section:Channel Estimation}, we discuss how the required channel parameters can be estimated using the RSSI feedback values. Then, in Section \ref{section:Results and Discussion}, we provide useful insights on the results through simulations. Section \ref{section:Conclusion} concludes the paper.

\section{System Model and Problem Setup}\label{Section:System model}
We consider the WET between an ET consisting of $K$ transmit antennas and two ERs having a single antenna each, over the wireless medium. For simplicity and the clarity of the analysis, it is assumed that $K=2$. The ET utilizes $M \leq K$ beams for the WET, and the transmit signal can be written as
$$\vec{x} = \frac{\sqrt{P}}{M}\sum_{m=1}^{M} \vec{b}_m s,$$
where $\vec{b}_m=\sqrt{\frac{1}{2}}\sqparen{1 \  \e{j\theta_m}}^\top$ is the beamforming vector of the $m$-th beam, and $s$ denotes the transmit symbol having unit power. 
It is assumed that the maximum sum-power constraint at the ET is $P>0$, $\ie$, $\mathrm E(\|\vec{x}\|^2)=\mathrm{tr}\paren{\vec{C_{xx}}} \leq P$, where $\vec{C_{xx}} = \mathrm{E(\vec{xx^\dagger})}$ is the transmit covariance matrix. The ET transmits using EGT beamforming similar to \cite{our_rssi,our_tsp}, where the ET equally splits the power among all transmit antennas, and focuses on adding the transmitted signals coherently at the ERs \cite{why_mgt}. In practice, each transmit antenna has its own power amplifier, which operates properly only when the transmit power is below a pre-designed threshold. EGT can guarantee that the transmit power will not exceed this threshold, unlike with maximum ratio transmit (MRT) beamforming, where the transmit power in some antennas may theoretically exceed these threshold values. Because of this reason, EGT beamforming is still a preferred method in practice.

Let $\vec{h}_{i} = \sqparen{ |h_{i1}| \e{j\delta_{i1}} \ \  |h_{i2}| \e{j\delta_{i2}}}^\top $ be the random complex multiple-input-single-output (MISO) channel vector between the ET and the $i$-th ER. $\vec{h}_{i}$ includes the effect of both path-loss and multipath fading between the ET and the $i$-th ER, and it is considered to be i.i.d. with an arbitrary distribution. We assume a quasi-static block fading channel where the channel is assumed to be fixed over a given block of transmission. When both energy beams are being transmitted, the RSSI value at the $i$-th ER can be written as
\begin{eqnarray}
R_{i}\paren{\theta_1,\theta_2} =   \vec{h}_{i}^{\dagger} \vec{C_{xx}} \vec{h}_{i}.
\label{RSSI}
\end{eqnarray}

It is well known that CSI plays a vital role in beamforming. Therefore, the WET process consists of two stages. Firstly, we have the training stage that the ET uses for channel learning. Then, the knowledge on the channel is used to set the beamformer in the second stage where the actual WET happens. We call this the wireless power beamforming (WPB) stage.  As we are focusing on applications with low complex ERs having limited processing capabilities, the well known method of estimating the channel at the ER, may become infeasible. Channel estimation involves analog to digital conversion and baseband processing, which require significant energy. Therefore, we focus on the low complex RSSI based channel learning technique introduced in \cite{our_rssi,our_tsp}.  In this scheme, the ER will feed back a set of RSSI values back to the ET, and these values are utilized for channel learning.  
We will elaborate the channel estimation process in detail in Section \ref{Section:Channel Estimation}. 


We focus on maximizing the average energy transfer in the WPB stage, in each transmission block. However, we ensure that a minimum harvested energy criterion for each ER is fulfilled, so that they can stay operational. In this way, we guarantee some fairness in the WET while being conscious on increasing the overall performance of the WET. More specifically, we set $\theta_1$ and $\theta_2$ in $\vec{b}_1$ and $\vec{b}_2$, respectively, to maximize the average sum energy transfer in the WPB stage. We formulate our optimization problem as follows.
\begin{equation}\label{eqn: opt problem}
\begin{aligned}
& \underset{\theta_1, \theta_2} {\text{maximize}}
& & \sum_{i=1}^{2} {R}_{i}\paren{\theta_1,\theta_2} \\
& \text{subject to}
& & {R}_i\paren{\theta_1,\theta_2} \geq \rho_{i}, \; i = 1,2.\\
&&& \mathrm{tr}\paren{\vec{C_{xx}}} \leq P
\end{aligned}
\end{equation}
This is a non-convex optimization problem. Also, it is not hard to see that the solution to this optimization problem will be a function of various channel parameters such as channel phase and amplitude. We should stress that perfect CSI is unavailable at the ET. Therefore, any analytical solution should only be a function of channel parameters that can be estimated using the RSSI feedback values. Obviously, some parameters cannot be estimated by only utilizing RSSI feedback values, and this can be considered a major bottleneck.

%
%

\section{Solving the Optimization Problem} \label{Section:Solving Optimization}
We start the analysis by obtaining a mathematically tractable expression for the objective function of the optimization problem of interest. To this end, by substituting for the channel covariance matrix, 
   %
\eqref{RSSI} can be further simplified, and the RSSI in a given transmission block at the $i$-th ER can be written as 
\begin{eqnarray}
{R}_{i}\paren{\theta_1,\theta_2} = \alpha_{i} + \beta_{i}\cos\paren{\theta_1-\theta_2} +\gamma_{i}g\paren{\theta_1,\theta_2,\phi_{i}},
\label{avg_rssi_single}
\end{eqnarray}
where $$g\paren{\theta_1,\theta_2,\phi_i} = \cos\paren{\theta_1+\phi_i}+\cos\paren{\theta_2+\phi_i},$$  $\alpha_{i}=\frac{P}{8}\paren{|2h_{i1}|^{2}+|h_{i2}|^2}$,  $\beta_{i}=\frac{P}{8}|h_{i2}|^2$, $\gamma_{i}=\frac{P}{4}|h_{i1}||h_{i2}|$ and $\phi_{i} = \delta_{i2}-\delta_{i1}$ (the phase difference between $h_{i2}$ and $h_{i1}$).
This can be utilized to obtain an expression for the objective function, which is formally stated in the following lemma. 
\begin{lemma} \label{Lemma: objective}
Let $k_{1} = \gamma_{1}\cos\paren{\phi_{1}}+\gamma_{2}\cos\paren{\phi_{2}}$ and $k_{2} = \gamma_{1}\sin\paren{\phi_{1}}+\gamma_{2}\sin\paren{\phi_{2}}$. The sum of the RSSI values at the two ERs is given by 
\begin{eqnarray}
{R}_{T}\paren{\theta_1,\theta_2}=\alpha_{T}+\beta_{T}\cos\paren{\theta_1-\theta_2}+\gamma_{T}g\paren{\theta_1,\theta_2,\phi_{T}},
\label{avg_rssi_total}
\end{eqnarray}
where
$\alpha_{T} = \alpha_{1} + \alpha_{2}$,
$\beta_{T} = \beta_{1} + \beta_{2}$,  
$\gamma_{T} = \sqrt{k_{1}^{2}+k_{2}^{2}}$ and 
$\tan(\phi_{T})=\frac{k_{2}}{k_{1}}$.
\end{lemma}
\begin{IEEEproof}
We have the following:
\begin{eqnarray*}
{R}_{T}\paren{\theta_1,\theta_2} = \alpha_{T}+\beta_{T}\cos\paren{\theta_1-\theta_2}+\sum_{i=1}^{2}\gamma_{i}g\paren{\theta_1,\theta_2,\phi_{i}},\\
g\paren{\theta_1,\theta_2,\phi_{i}} = \sum_{m=1}^2\sqparen{\cos(\theta_m)\cos(\phi_i) - \sin(\theta_m)\sin(\phi_i)}.
\end{eqnarray*}
Thus, 
\begin{multline*}
 \sum_{i=1}^{2}\gamma_{i}g\paren{\theta_1,\theta_2,\phi_{i}}= \\ \sqrt{k_{1}^{2}+k_{2}^{2}}\sqparen{\sum_{m=1}^2\sqparen{\cos(\theta_m)\cos(\phi_T) - \sin(\theta_m)\sin(\phi_T)}},
\end{multline*}
which completes the proof.
\end{IEEEproof}
The form obtained for the objective function particularly facilitates obtaining a solution for the optimization problem, which we will see next.

\subsection{Special case: $\rho_{1} = \rho_{2} = 0$ }
Firstly, we will consider solving an unconstrained optimization problem, $\ie$, we consider $\rho_{1} = \rho_{2} = 0$. This means the ET focuses on maximizing the harvested energy without considering fairness in the network. We use this case to obtain some interesting insights, and also, as an auxiliary result in solving the general optimization problem. 
\begin{theorem} \label{thm:unconstrained}
When $\rho_{1}=\rho_{2}=0$, the average energy transfer in the WPB stage is maximized when $\theta_1=\theta_2= -\phi_T $.
\end{theorem}
\begin{IEEEproof} It can be easily seen that setting $\theta_1=\theta_2= -\phi_T $ maximizes all three cosine terms in the objective function given in Lemma \ref{Lemma: objective}, which in turn maximizes the objective function, completing the proof. 
\end{IEEEproof}

It is interesting to note that using the same vector for both beams is optimal, which implies that transmitting one beam is always optimal if there are no constraints on the harvested energy of the individual ERs. To this end, when the constraints are relaxed, we end up with a convex optimization problem, which is easy to solve.  Intuitively, we can expect the same single beam optimality when there are more than two ERs, and the idea is to allocate all the power to a single beam, and transmit it in the direction of a centroid associated with the channels between the ET and ERs. Finding this centroid will depend on what sort of CSI can be obtained at the ET. In fact, the result goes in line with \cite{rank_one}, where the optimality of a single beam is claimed for a scenario of broadcasting.

When a single beam is being transmitted, the result in Lemma \ref{Lemma: objective} can be further simplified as    
\begin{eqnarray}
{R}_{T}\paren{\theta} &= \alpha^{\prime}_{T} + \gamma^{\prime}_{T}cos(\theta+\phi_{T}),
\label{rssi_onebeam}
\end{eqnarray}
where $\alpha^{\prime}_{T} = \alpha_{T} + \beta_{T}, \gamma^{\prime}_{T} = 2\gamma_{T}$. Replacing index $T$ with index $i$ in all the notations in \eqref{rssi_onebeam} gives us ${R}_{i}\paren{\theta}$, which is the RSSI in the WPB stage for ER $i$ given a single beam is transmitted. To set the beamforming vector that maximizes ${R}_{T}\paren{\theta}$, we should be able to estimate $\phi_T$, and this will be studied in Section \ref{Section:Channel Estimation}.  Next, we will consider the general constrained optimization problem.

\subsection{Solving the constrained optimization problem}

We will now consider solving the two-dimensional constrained optimization problem of interest. As mentioned earlier, this in general is a non-convex optimization problem, and turns out to be convex if the constraints on the harvested energy are omitted. Therefore, obtaining an analytical solution for the general case is difficult. Also as highlighted previously, the solution should be a function of channel parameters that can be estimated at the ET. For an example, in Theorem \ref{thm:unconstrained}, the solution was $\phi_{T}$, and this value has to be estimated to set the beamforming vector. Therefore, we focus on a suboptimal solution. However, although being suboptimal, it can be practically implemented in a system with low complex ERs. To this end, we will assume that only one beam is transmitted in the WPB stage, which simplifies the problem to a single dimensional optimization problem. This assumption will in deed lead to loss of optimality as there will be cases where transmitting two beams will give better performance compared to transmitting a single beam. However, we will show that if the channels between the ET and the ERs satisfy a certain condition, transmitting a single beam will be optimal, and this condition is formally presented through the following lemma.
\begin{lemma}
If $|\phi_{1}-\phi_{2}|\leq\frac{\pi}{2}$, the average energy transfer in the WPB stage is maximized when $\theta_1=\theta_2$, that is, by transmitting a single energy beam. 
\label{lemma2}
\end{lemma}
\begin{IEEEproof}
Let $\tilde{\theta}_1 = \frac{\theta_1+\theta_2}{\sqrt{2}}$ and $\tilde{\theta}_2=\frac{\theta_2-\theta_1}{\sqrt{2}}$, which leads to
\begin{multline*}
{R}_{i}\paren{\tilde{\theta}_1,\tilde{\theta}_2} = \alpha_{i} + \beta_{i}\cos\paren{\sqrt{2}\tilde{\theta}_2} \\+2\gamma_{i}\cos(\frac{\tilde{\theta}_2}{\sqrt{2}})\cos\paren{\frac{\tilde{\theta}_1}{\sqrt{2}}+\phi_{i}} 
\end{multline*}
for $i \in \brparen{1,2,T}$. Without loss of generality, let us assume $-\phi_{2}\geq -\phi_{1}$. When $|\phi_{1}-\phi_{2}|\leq\frac{\pi}{2}$, that is, when the phase difference of the two channels is an acute angle, it can be shown that $-\phi_{1}\leq-\phi_{T}\leq -\phi_{2}$.
We can also show that $$\frac{-2\phi_1}{\sqrt{2}} \leq \tilde{\theta}_1 \leq \frac{-2\phi_2}{\sqrt{2}}$$ and $$\frac{-\pi}{2\sqrt{2}} \leq \tilde{\theta}_2 \leq \frac{-\pi}{2\sqrt{2}},$$ which imply $\left|\frac{\tilde{\theta}_1}{\sqrt{2}}+\phi_{i}\right| \leq \frac{\pi}{2}$ and $\left|\tilde{\theta}_2\right|\leq \frac{\pi}{2\sqrt{2}}$  for $i = \brparen{1,2,T}$. By using these inequalities, we get 

\[
{R}_{i}\paren{\tilde{\theta}_1,\tilde{\theta}_2} = \alpha_{i} + \beta_{i}cos\paren{\sqrt{2}\tilde{\theta}_2} +2\tilde{\gamma}_{i}\cos\paren{\frac{\tilde{\theta}_2}{\sqrt{2}}} , \]
where
$\tilde{\gamma}_{i} = \cos\paren{\frac{\tilde{\theta}_1}{\sqrt{2}}+\phi_{i}} \geq 0$ for $i = \brparen{1,2,T}$. 

Assume $\exists$ a point $\paren{\tilde{\theta}_{1}^\tau,\tilde{\theta}_{2}^\tau}$, such that $\tilde{\theta}_{2}^\tau \neq 0$, and satisfies both constraints. 
If we consider a point $\paren{\tilde{\theta}_{1}^\tau,\tilde{\theta}_{2}^\tau - \epsilon}$, we will have $${R}_i\paren{\tilde{\theta}_{1}^\tau,\tilde{\theta}_{2}^\tau - \epsilon} > {R}_i\paren{\tilde{\theta}_{1}^\tau,\tilde{\theta}_{2}^\tau}\geq \rho_{i}$$ for $i \in \brparen{1,2}$, and $${R}_T\paren{\tilde{\theta}_{1}^\tau,\tilde{\theta}_{2}^\tau - \epsilon} > {R}_T\paren{\tilde{\theta}_{1}^\tau,\tilde{\theta}_{2}^\tau}.$$  
Therefore, ${R}_{T}\paren{\tilde{\theta}_{1},\tilde{\theta}_{2}}$ is maximized when $\tilde{\theta}_2=0$, and both constraints are satisfied at this point. This implies $\theta_1 = \theta_2$, completing the proof. 
%
\end{IEEEproof}

When this sufficient condition holds, we can obtain an analytical solution to the optimization problem of interest, and this solution is formally stated in the following theorem.
\begin{theorem}\label{thorem2}
Let $\omega_{1} = \left|\cos^{-1}\paren{\frac{\rho_{1} - \alpha^{\prime}_{1}}{\gamma^{\prime}_{1}}}\right|$ and
$\omega_{2} = \left|\cos^{-1}\paren{\frac{\rho_{2} - \alpha^{\prime}_{2}}{\gamma^{\prime}_{2}}}\right|$. If $|\phi_{1}-\phi_{2}|\leq\frac{\pi}{2}$, the average total energy transfer in the WPB stage is maximized when $\theta_1=\theta_2= \theta^\star $,
where
\[
    \theta^{*}= 
\begin{cases}
    -\phi_{T},& \text{if } -\phi_{T} \in \sqparen{\psi_{1},\psi_{2}}\\
    \arg \max \{{R}_T \paren{\psi_{1}},{R}_T\paren{\psi_{2}}\}              & \text{otherwise}
\end{cases},
\]
in which  
 $$[\psi_{1},\psi_{2}] \in [-\phi_{1}-\omega_{1},-\phi_{1}+\omega_{1}]\cap [-\phi_{2}-\omega_{2},-\phi_{2}+\omega_{2}] .$$
\end{theorem}
\begin{IEEEproof}
Firstly, we obtain intervals of $\theta$ that satisfy the individual energy constraints. To this end, we have
 $$\cos(\theta+\phi_{1}) \geq \frac{\rho_{1} - \alpha^{\prime}_{1}}{\gamma^{\prime}_{1}}$$ and $$\cos(\theta+\phi_{2}) \geq \frac{\rho_{2} - \alpha_{2}^{'}}{\gamma^{\prime}_{2}}.$$ 
By solving the above inequalities, we get
$[-\phi_{1}-\omega_{1},-\phi_{1}+\omega_{1}]$ and $ [-\phi_{2}-\omega_{2},-\phi_{2}+\omega_{2}]$, which are convex sets over $\theta$. The intersection between these two regions gives us $[\psi_{1},\psi_{2}]$, which is again convex over $\theta$. If the maximum of the objective function lies in this feasible region, it is achievable, and from Theorem \ref{thm:unconstrained}, we know that this point can be achieved by setting $\theta^\star= -\phi_T$. 

Looking at the case of this maximum point being infeasible, it is not hard to see according to \eqref{rssi_onebeam} that when transmitting a single beam, the objective function takes the form of a generic one dimensional cosine function. In a given convex region, a cosine function can take a maximum$\slash$minimum or it can be strictly increasing$\slash$decreasing. Therefore, if the global maximum point of the objective function is not in the feasible region, the highest value that the objective function can achieve should be at one of the boundary points in the region, $\ie$, $\omega_1$ or $\omega_2$, which completes the proof. 
\end{IEEEproof}

In practice, there can be WET requirements where there is only a single priority ER with a minimum power constraint. We can directly use the result in Theorem \ref{thorem2} for scenarios like that, and the extension is presented through the following corollary. 
\begin{corollary}
Let $\rho_1 > 0$, $\rho_2 = 0$ and $\omega_{1} = \left|\cos^{-1}\paren{\frac{\rho_{1} - \alpha^{\prime}_{1}}{\gamma^{\prime}_{1}}}\right|$.
If $|\phi_{1}-\phi_{2}|\leq\frac{\pi}{2}$, the average energy transfer in the WPB stage is maximized when $\theta_1=\theta_2= \theta^\star $,
where
\[
    \theta^{*}= 
\begin{cases}
    -\phi_{T},& \text{if } -\phi_{T} \in \sqparen{\psi_{1},\psi_{2}}\\
    \arg \max \{{R}_T \paren{\psi_{1}},{R}_T\paren{\psi_{2}}\}              & \text{otherwise}
\end{cases},
\]
in which  
 $[\psi_{1},\psi_{2}] \in [-\phi_{1}-\omega_{1},-\phi_{1}+\omega_{1}]$.
 \label{corollary1}
\end{corollary}

We will set our beamforming vector in the WPB stage according to Theorem \ref{thorem2}. This will be the optimal selection of the beamforming vector if $|\phi_{1}-\phi_{2}|\leq\frac{\pi}{2}$, and suboptimal otherwise. Note that this is a sufficient condition, and not a necessary condition, which means the optimality may in fact hold for a larger region. We will get more insights on this region through simulations in Section \ref{section:Results and Discussion}. If the phase values are uniformly distributed, the sufficient condition guarantees optimality for 50$\%$ of all instances on the average.

Next, we will have to look at the feasibility of the solution. Some of our suboptimal solutions can be infeasible, that is, we might not be able to satisfy the individual energy constraints in a given block of transmission, due to the channels being poor. To this end, we propose Algorithm \ref{euclid} as a suitable transmit algorithm for the WPB stage.
\begin{algorithm}
\caption{Transmit Algorithm}\label{euclid}
\begin{algorithmic}[1]
\State $\rho_1 \gets \textit{constraint 1}$
\State $\rho_2 \gets \textit{constraint 2}$
\If {$[\psi_{1},\psi_{2}] \neq \emptyset$} 
	\State\text{use} \textit{Theorem \ref{thorem2}} \text{and transmit}
\Else 
     \State $\min\paren{\rho_1,\rho_2} \gets 0$
     \If {$[\psi_{1},\psi_{2}] \neq \emptyset$} 
     	\State\text{use} \textit{Corollary \ref{corollary1}} \text{and transmit}
     \Else 
     	\State $\rho_1 \gets \textit{constraint 1} \text{, } \rho_2 \gets \textit{constraint 2}$
          \State $\max\paren{\rho_1,\rho_2} \gets 0$
           \If {$[\psi_{1},\psi_{2}] \neq \emptyset$} 
               	\State\text{use} \textit{Corollary \ref{corollary1}} \text{and transmit}
               \Else 
		        \State $\rho_1 \gets 0 \text{, } \rho_2 \gets 0$
                \State\text{use} \textit{Theorem \ref{thm:unconstrained}} \text{and transmit}
               \EndIf
     \EndIf
\EndIf

\end{algorithmic}
\end{algorithm}
According to the algorithm, we first try to satisfy both ERs. If this step is infeasible, we try to satisfy only the ER with the higher energy requirement. If the second step is infeasible as well, we try to satisfy only the ER with the lower energy requirement, and if it is still infeasible, we neglect both constraints and try to maximize the average sum energy transfer.   

\subsection{Extension for multiple ERs}
From the results in Theorem \ref{thorem2}, we can get interesting insights into a scenario with multiple ERs. In a scenario with multiple ERs, we are more likely to have two ERs with $|\phi_{1}-\phi_{2}|\leq\frac{\pi}{2}$. This means, we can always find a pair of ERs satisfying the sufficient condition in Theorem \ref{thorem2} if the number of ERs is large enough. Hence, we can introduce a pairwise scheduling policy without any loss of optimality. Moreover, we can schedule two ERs which are closest to each other in terms of channel phase, and the energy transfer will be much efficient than trying to schedule two ERs that are distant from each other (in terms of channel phase). However, note that according to this scheduling policy, the constraints of the ERs not being scheduled may be violated in a given transmission block. Therefore, some sort of time averaging should be introduced for the constraints, and they should be set appropriately such that over time, the required energy requirement is satisfied. Since the channel phase values are i.i.d., each ER has equal chance of being scheduled. More insights on this scheduling policy will be presented in Section \ref{section:Results and Discussion} with appropriate simulations.    

We have already claimed that the main reason for going for this suboptimal method is the lack of perfect CSI at the ET. However, even to implement this proposed algorithm, the ET has to know $\alpha_i^\prime$, $\gamma_i^\prime$, $\phi_i$ for $i=1,2$, and $\phi_T$. The estimation of these seven parameters will be discussed in the next section.

\section{Training and Channel Estimation}\label{Section:Channel Estimation}
We will first focus on obtaining the RSSI feedback values, or more specifically, the training stage of the proposed methodology. The training stage is similar to the one proposed in \cite{our_rssi,our_tsp}. In the training stage, the ET transmits a single beam by sequentially changing the direction of the beam (beamforming vector) according to a pre-defined codebook of size $N$. To this end, the codebook will take the form of $\vec{B} = \brparen{\vec{b}_1^1, \ldots,\vec{b}_1^N}$, which includes $N$ beamforming vectors where $\vec{b}_1^j$ takes the form $\sqparen{1\ \  \e{j\theta_{j}}}^\top$ and $\theta_{j} = \frac{2(j-1)\pi}{N}$ for $j \in \brparen{1,\ldots,N}$. The corresponding RSSI value for each beamforming vector is fed back by the ERs to the ET. That is, each ER will feed back $N$ RSSI values. In \cite{our_rssi,our_tsp}, only $\phi_i$ had to be estimated, and it is shown that this codebook allows the estimator of $\phi_i$ to achieve the Cramer-Rao-Lower-Bound (CRLB). It can be shown that this is true for other parameters that we need to estimate as well, thus, we use the same structure.

The instantaneous RSSI value at the $i$-th ER corresponding to the $j$-th beamforming vector in the codebook can be written as,
\begin{eqnarray}
R_{i}\paren{\theta_j} &= \alpha^{\prime}_{i} + \gamma^{\prime}_{i}\cos(\theta_j+\phi_{i}) + z_i,
\label{rssi_with_noise}
\end{eqnarray}
where $ i =1,2$. Replacing index $i$ with $T$ in \eqref{rssi_with_noise} gives us an expression for $R_T\paren{\theta_j}$, and $R_T\paren{\theta_j}=R_1\paren{\theta_j}+R_2\paren{\theta_j}$. Although we have assumed a quasistatic block-fading channel, the RSSI value will change from one measurement to the other due to the effect of noise. We have used random variable $z_i$ to represent this effect. To this end, $z_i$ captures the effect of all noise related to the measurement process such as noise in the channel, circuit, antenna matching network and rectifier, and we assume the
random variables to be i.i.d. additive Gaussian with zero
mean and variance $\sigma^2$. This means, we also assume that in a given
transmission block, the randomness in \eqref{rssi_with_noise} is caused only by
$z_i$. Based on these assumptions, estimating the channel parameters becomes a classical parameter estimation problem \cite{est_kay}. A maximum likelihood estimation can be carried out by utilizing 
\begin{eqnarray}
\mathrm E_i \defeq \sum_{j=1}^{N}\sqparen{R_{i}\paren{\theta_j} - \paren{\alpha^{\prime}_{i} +\gamma^{\prime}_{i}\cos(\theta_j+\phi_{i})}}^2
\label{sum_of_res}
\end{eqnarray}
for $ i \in \brparen{1,2,T}$. Moreover, differentiating $\mathrm E_i$ with respect to  $\alpha_{i}^\prime$, $\gamma_{i}^\prime$ and $\phi_{i}$, respectively, and equating to zero gives us the required estimates, which are stated in the following lemma.
\begin{lemma}\label{estimates}
For a sample of $ N $ i.i.d. RSSI observations, the maximum likelihood estimates of $\alpha_{i}^\prime$, $\gamma_{i}^\prime$ and $\phi_{i}$ are given by,
	\begin{gather}
	\hat{\phi}_{i} = \tan^{-1}\paren{\frac{\displaystyle  -\sum_{j=1}^{N} R_{i}\paren{\theta_j}\sin\theta_{j} }           
		{\displaystyle \sum_{j=1}^{N} R_{i}\paren{\theta_j} \cos\theta_{j}}	
	},
	\label{estimation_phi}
	\end{gather}

	\begin{gather}
	\hat{\alpha}_{i}^\prime = \frac{\displaystyle  \sum_{j=1}^{N} R_{i}\paren{\theta_j}}          
		{N}	
	,
	\label{estimation_alpha_i}
	\end{gather}
		
	\begin{gather}
		\hat{\gamma}_{i}^\prime = \frac{\displaystyle  2\sum_{j=1}^{N} R_{i}\paren{\theta_j}cos(\theta_j+\phi_i)}           
			{N}	
		,
		\label{estimation_gamma_i}
	\end{gather}
where $ \theta_{j} = \frac{2(j-1)\pi}{N}$ and $i=1,2$.
Expression for the estimate of $\phi_T$ can be obtained by replacing index $i$ in \eqref{estimation_phi} with $T$. 
\end{lemma}
\begin{IEEEproof}
Differentiating $\mathrm E_i$ with respect to  $\alpha_{i}^\prime$, $\gamma_{i}^\prime$ and $\phi_{i}$, respectively, equating to zero, and using $\sum_{j=1}^{N} \sin(\theta_{j}+\phi_{i})=\sum_{j=1}^{N} \sin{[2(\theta_{j}+\phi_{i})]}=$ $\sum_{j=1}^{N} \cos(\theta_{j}+\phi_{i})=\sum_{j=1}^{N} \cos{[2(\theta_{j}+\phi_{i})]}= $ $\sum_{j=1}^{N} \cos(\theta_{j}+\phi_{i})=\cos{[2(\theta_{j}+\phi_{i})]}=0 $, for simplification completes the proof \cite{trig_table}. 
\end{IEEEproof}	

The obtained results for the estimates are remarkably simple, can be implemented easily at the ET, and require minimal processing. However, in order to estimate $\gamma^\prime_i$, we have to use estimates of $\phi_i$ obtained from \eqref{estimation_phi}, thus, $\phi_i$ has to be estimated first. The ambiguity in $\phi_i$ can be resolved using a similar method proposed in \cite{our_tsp}. We should note that we cannot estimate all required parameters in \eqref{avg_rssi_single} and \eqref{avg_rssi_total} using RSSI based estimation. To be particular, we cannot estimate $\beta_i$ for $i=1,2$ and $\beta_T$. Therefore, we had to make a single beam assumption which nullifies the $\beta_i$ terms in the objective function. Now, we have all required variables estimated to implement the transmit algorithm in Section \ref{Section:Solving Optimization}.

\section{Simulation Results and Discussion} \label{section:Results and Discussion}
\begin{figure}[t]
	\centering{\includegraphics[scale=0.5]{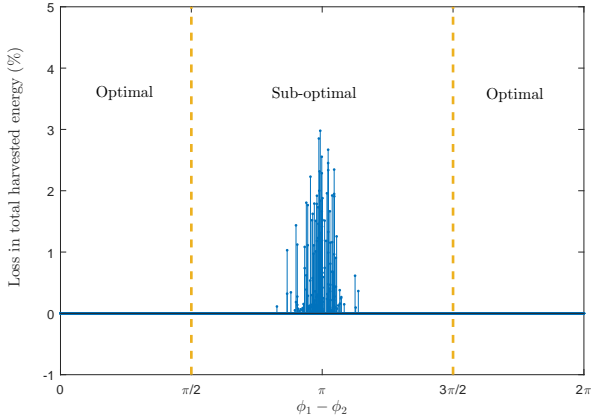}} 
	\caption{The optimality and sub-optimality regions when transmitting a single energy beam.}	 
	\label{Fig: optsub} \vspace{-0.3cm}
\end{figure}

In this section, we use numerical evaluations and simulation results to provide further insights. Firstly, we present the simulation result illustrated in Fig. \ref{Fig: optsub}. In Lemma \ref{lemma2}, we obtained a sufficient condition for the optimality of transmitting a single beam, $\ie$, we showed that when $|\phi_{1}-\phi_{2}|\leq\frac{\pi}{2}$, the average energy transfer in the WPB stage is maximized when $\theta_1=\theta_2$. However, due to not having perfect CSI  at the ET, we resorted to the suboptimal solution of transmitting a single beam even when $|\phi_{1}-\phi_{2}|>\frac{\pi}{2}$. To this end, Fig. \ref{Fig: optsub} illustrates the performance of transmitting a single beam against different values of $|\phi_{1}-\phi_{2}|$. In the simulation, we have generated the channel coefficients randomly and uniformly between the amplitudes of 0.1 and 1, and channel phase differences of $0$ and $2\pi$. We have considered 10000 channel realizations, and for each realization, we have considered 100 different possible combinations of $\rho_1$ and $\rho_2$.  We can see that when $|\phi_{1}-\phi_{2}|\leq\frac{\pi}{2}$, we get the optimal solution by transmitting a single beam. However, the interesting thing to note here is that the sub-optimality arises only when $|\phi_{1}-\phi_{2}|$ is close to $\pi$. This means the optimality condition is true for a much larger region. Also, by observing the solid line at zero error, we can see that even when $|\phi_{1}-\phi_{2}|$ is close to $\pi$, we get optimality for some combinations of $\rho_1$ and $\rho_2$. 

\begin{figure}[t]
	\centering{\includegraphics[scale=0.5]{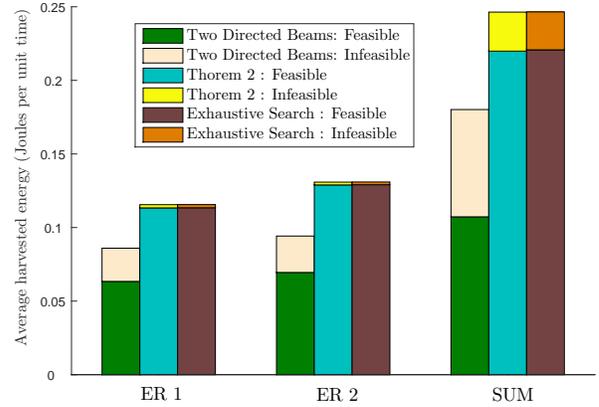}} 
	\caption{The average harvested energy at each ER for different beamforming methods.}	 
	\label{Fig: feasible} \vspace{-0.3cm}
\end{figure}

For the same simulation, we have illustrated the average harvested energy at each ER, and the total average harvested energy, in Fig. \ref{Fig: feasible}. It can be seen that the loss of optimality is negligibly small. After the training, one natural alternate option is to transmit two beams directed at each ER, $\ie$, by setting $\theta_1 = -\hat{\phi}_1$ and $\theta_2 = -\hat{\phi}_2$. The figure shows that our proposed method achieves a significant gain over this method. It should be noted that there can be scenarios where the ET is unable satisfy the minimum energy criterion at each ER. To this end, in Section \ref{Section:Solving Optimization}, we have proposed Algorithm 1, to continue WET by relaxing constraints. Fig \ref{Fig: feasible} also illustrates the performance of this algorithm, as it shows what is the harvested energy when the constraints are not satisfied. In the sum, the infeasible part contains the amount of harvested energy when at least one individual constraint is not satisfied, where as the ER-wise infeasible part contains the amount of harvested energy when its own energy constraint is not satisfied. When having more than two ERs, the ET is more likely to find a pair of ERs which satisfy the sufficient condition. Also, since the ERs are more likely to be closer to each other in terms of phase, the pairwise sum harvested energy should increase. Satisfying the constraints will be comparatively  easier as well when the two ERs are closer to each other, and hence, the infeasible region is expected to shrink when the number of ERs increases. An extension that stems on the idea presented in this paper on scheduling ERs that are closer to each other in terms of their channel phase can be found in \cite{samith_wiopt}.

\section{Conclusions} \label{section:Conclusion}
This paper has focused on transferring energy to a pair of low complex energy receivers (ER) over the wireless medium by using energy beamforming. The channel learning has been done at the energy transmitter (ET) by only utilizing RSSI values that are fed back from the ERs to the ET. Starting with two ERs, selecting the beamformer that maximizes the total average energy transfer between the ET and the ERs, while satisfying a minimum harvested energy criterion at each ER, has been studied. Analytical solutions to the optimization problem have been presented, together with sufficient conditions for the optimality, and algorithms for practical implementation. The proposed policy, although being suboptimal, is suitable for ERs having tight energy constraints and hardware limitations, and it has been shown that the optimality gap is negligibly small. Insights on the extension of results to an arbitrary number of ERs have been presented through a pairwise scheduling policy, that does not violate optimality. Performance gains have been highlighted using numerical evaluations and simulations.

\balance 

\footnotesize {\bibliography{bibfile}}

\begin{thebibliography}{10}

\bibitem{survey}
X.~Lu, P.~Wang, D.~Niyato, D.~I. Kim, and Z.~Han, ``Wireless networks with {RF}
  energy harvesting: A contemporary survey,'' {\em IEEE Commun. Surveys and
  Tutorials}, vol.~17, pp.~757--789, Nov. 2015.

\bibitem{power1}
R.~Zhang and C.~K. Ho, ``{MIMO} broadcasting for simultaneous wireless
  information and power transfer,'' {\em IEEE Trans. Wireless Commun.},
  vol.~12, pp.~1989--2001, May 2013.

\bibitem{multi_ant}
X.~Chen, Z.~Zhang, H.~h.~Chen, and H.~Zhang, ``Enhancing wireless information
  and power transfer by exploiting multi-antenna techniques,'' {\em IEEE
  Commun. Magazine}, vol.~53, pp.~133--141, Apr. 2015.

\bibitem{multi_mimo}
J.~Xu, S.~Bi, and R.~Zhang, ``Multiuser {MIMO} wireless energy transfer with
  coexisting opportunistic communication,'' {\em IEEE Wireless Commun.
  Letters}, vol.~4, pp.~273--276, Jun. 2015.

\bibitem{6623072}
X.~Chen, X.~Wang, and X.~Chen, ``Energy-efficient optimization for wireless
  information and power transfer in large-scale mimo systems employing energy
  beamforming,'' {\em IEEE Wireless Commun. Letters}, vol.~2, pp.~667--670,
  Dec. 2013.

\bibitem{7060673}
W.~Huang, H.~Chen, Y.~Li, and B.~Vucetic, ``On the performance of multi-antenna
  wireless-powered communications with energy beamforming,'' {\em IEEE Trans.
  Veh. Technol.}, vol.~65, pp.~1801--1808, Mar. 2016.

\bibitem{6568923}
X.~Chen, C.~Yuen, and Z.~Zhang, ``Wireless energy and information transfer
  tradeoff for limited-feedback multiantenna systems with energy beamforming,''
  {\em IEEE Trans. Veh. Technol.}, vol.~63, pp.~407--412, Jan. 2014.

\bibitem{Tse02}
P.~Viswanath, D.~Tse, and R.~Laroia, ``Opportunistic beamforming using dumb
  antennas,'' {\em IEEE Trans. Inf. Theory}, vol.~48, pp.~1277--1294, Jun.
  2002.

\bibitem{hassibi}
M.~Sharif and B.~Hassibi, ``On the capacity of {MIMO} broadcast channels with
  partial side information,'' {\em IEEE Trans. Inf. Theory}, vol.~51,
  pp.~506--522, Feb. 2005.

\bibitem{ts_on_the_outage}
T.~Samarasinghe, H.~Inaltekin, and J.~S. Evans, ``On the outage capacity of
  opportunistic beamforming with random user locations,'' {\em IEEE Trans.
  Commun.}, vol.~62, pp.~3015--3026, Aug. 2014.

\bibitem{ts_modandanalysis}
T.~Samarasinghe, H.~Inaltekin, and J.~S. Evans, ``Modeling and analysis of
  opportunistic beamforming for poisson wireless networks,'' {\em IEEE Trans.
  Wireless Commun.}, vol.~15, pp.~3732--3745, May 2016.

\bibitem{one_bit}
J.~Xu and R.~Zhang, ``Energy beamforming with one-bit feedback,'' {\em IEEE
  Trans. Signal Process.}, vol.~62, pp.~5370--5381, Oct. 2014.

\bibitem{recip3}
Y.~Zeng and R.~Zhang, ``Optimized training for net energy maximization in
  multi-antenna wireless energy transfer over frequency-selective channel,''
  {\em IEEE Trans. Commun.}, vol.~63, pp.~2360--2373, Jun. 2015.

\bibitem{rssi_work}
S.~Lakshmanan, K.~Sundaresan, S.~Rangarajan, and R.~Sivakumar, ``Practical
  beamforming based on {RSSI} measurements using off-the-shelf wireless
  clients,'' in {\em Proc. Internet measurement conference}, pp.~410--416, Nov.
  2009.

\bibitem{our_rssi}
S.~Abeywickrama, T.~Samarasinghe, and C.~K. Ho, ``Wireless energy beamforming
  using signal strength feedback,'' in {\em Proc. IEEE Global
  Telecommunications Conference}, pp.~1 -- 6, Dec. 2016.

\bibitem{our_tsp}
S.~Abeywickrama, T.~Samarasinghe, C.~K. Ho, and C.~Yuen, ``Wireless energy
  beamforming using received signal strength indicator feedback,'' {\em IEEE
  Trans. Signal Process.}, vol.~66, pp.~224--235, Jan. 2018.

\bibitem{why_mgt}
S.-H. Tsai, ``Equal gain transmission with antenna selection in {MIMO}
  communications,'' {\em IEEE Wireless Commun.}, vol.~10, pp.~1470--1479, May
  2011.

\bibitem{rank_one}
Y.~Huang and D.~P. Palomar, ``Rank-constrained separable semidefinite
  programming with applications to optimal beamforming,'' {\em IEEE Trans.
  Signal Process.}, vol.~58, pp.~664 -- 678, Feb. 2010.

\bibitem{est_kay}
S.~M. Kay, {\em Fundamentals of Statistical Signal Processing: Estimation
  Theory}.
\newblock Upper Saddle River, NJ, USA: Prentice-Hall, Inc., 1993.

\bibitem{trig_table}
I.~S. Gradshteyn and I.~M. Ryzhik, {\em Table of integrals, series, and
  products}.
\newblock Elsevier/Academic Press, Amsterdam, seventh~ed., 2007.

\bibitem{samith_wiopt}
S.~Abeywickrama, T.~Samarasinghe, C.~Yuen, and R.~Zhang, ``Cluster-based
  wireless energy transfer for low complex energy receivers,'' {\em To appear
  in International Symposium on Modeling and Optimization in Mobile, Ad Hoc and
  Wireless Networks}, 2018.

\end{thebibliography}
\end{document}